\documentclass[twocolumn,showpacs,amsfonts,aps,prl,floatfix,nofootinbib,float,superscriptaddress]{revtex4-1} 
\usepackage{graphicx}  % needed for figures
\usepackage{dcolumn}   % needed for some tables
\usepackage{bm}        % for math
\usepackage{amssymb}   % for math
\usepackage{amsmath}   % for math

\newcommand{\beq}{\begin{equation}}
\newcommand{\eeq}{\end{equation}}
\newcommand{\bea}{\begin{eqnarray}}
\newcommand{\eea}{\end{eqnarray}}
\newcommand{\beas}{\begin{eqnarray*}}
\newcommand{\eeas}{\end{eqnarray*}}

\newcommand{\Tint}[1]{{\hbox{$\sum$}\!\!\!\!\!\!\!\int\,}_{\!\!\!\!\raise-0.9ex\hbox{$\scriptstyle{#1}$}}}

\begin{document}
\widetext

\title{A novel Bayesian approach to spectral function reconstruction}

\author{Yannis Burnier, Alexander Rothkopf}
\affiliation{Albert Einstein Center for Fundamental Physics, University of Bern, Sidlerstrasse 5, CH-3012 Bern, Switzerland}

\date{\today}

\begin{abstract}
We present a novel approach to the inference of spectral functions from Euclidean time correlator data that makes close contact with modern Bayesian concepts. Our method differs significantly from the maximum entropy method (MEM). A new set of axioms is postulated for the prior probability, leading to an improved expression, which is devoid of the asymptotically flat directions present in the Shanon-Jaynes entropy. Hyperparameters are integrated out explicitly, liberating us from the Gaussian approximations underlying the evidence approach of the MEM. We present a realistic test of our method in the context of the non-perturbative extraction of the heavy quark potential. Based on hard-thermal-loop correlator mock data, we establish firm requirements in the number of data points and their accuracy for a successful extraction of the potential from lattice QCD. An improved potential estimation from previously investigated quenched lattice QCD correlators is provided.\vspace{-0.2cm}
\end{abstract}

\maketitle
The numerical solution of inverse problems is an active area of research with important applications in science and engineering. In the context of QCD physics, the estimation of spectral functions from Euclidean correlators is of particular interest. A reliable determination of ground and excited state properties of mesons and baryons at zero temperature \cite{Sasaki:2005ap} from non-perturbative Monte-Calo simulations (lattice QCD) e.g. represents an important bridge between field theory and experiment. At finite temperature, lattice spectral functions allow us to scrutinize the physics of the early universe by elucidating phenomena, such as heavy quarkonium melting \cite{Asakawa:2003re} and the transport properties \cite{Aarts:2007wj} of the quark-gluon plasma produced in relativistic heavy-ion collisions.

The most common approach to spectral function reconstruction deployed today, the maximum entropy method (MEM) \cite{Jarrell:1996}, is based on Bayesian inference. Nevertheless even after 20 years of application \cite{Asakawa:2000tr,Asakawa:2003re,Nickel:2006mm,Rothkopf:2011db}, the reliability of the MEM is still under discussion \cite{Gunnarsson:2010,Rothkopf:2011,Burnier:2013fca}. Here we introduce a novel Bayesian approach that addresses key issues affecting the MEM: \textit{slow convergence} of the underlying optimization task, high \textit{computational cost} for extended search spaces, \textit{scale dependence} in the prior functional and the \textit{Gaussian approximation} required in the hyperparameter estimation.

The Bayesian strategy \cite{Bishop:2007} relies on an application of the multiplication law for the joint probability distribution of the spectral function of the system under investigation $\rho$, the measured data $D$ and any other prior information $I$ 
\begin{align}
 P[\rho,D,I]\quad \Rightarrow \quad P[\rho|D,I]=\frac{P[D|\rho,I]P[\rho|I]}{P[D|I]}. \label{Eq:Bayes}
\end{align}
We specify in the likelihood probability $P[D|\rho,I]$ how the data is obtained, while the prior probability $P[\rho|I]$ encodes how prior information on $\rho$ itself enters the posterior $P[\rho|D,I]$. The maximum of $P[\rho|D,I]$ will ultimately provide us with a point estimate of $\rho$, which we refer to as the Bayesian solution to the inverse problem.

In the following, we aim at inverting the convolution
\begin{align}
 D(\tau)=\int d\omega  K(\omega,\tau)\rho(\omega),
\end{align}
which connects the spectral function $\rho(\omega)>0$ through a known kernel function $K(\tau,\omega)$ to the correlation function $D(\tau)$. In practice the correlator is estimated at $N_\tau$ points $D(\tau_i)=D_i$ from a sample of Gaussian distributed measurements. After discretization of the frequencies along $N_\omega$ points spaced by $\Delta\omega_l=\omega_{l+1}-\omega_l$ we can compute the corresponding data for each spectrum $\rho(\omega_l)=\rho_l$

\begin{align}
D^\rho_i=\sum_{l=1}^{N_\omega}\, \Delta\omega_l\, K_{il} \rho_l. \label{Eq:ConvolutionDiscr}
\end{align}
According to the Gaussian assumption, we use the quadratic distance 
\begin{align}
  &L=\frac{1}{2}\sum_{ij}(D_i-D^\rho_i)C^{-1}_{ij}(D_j-D^\rho_j),
\end{align}
to assign a probability to the data given a test spectral function. Here $C_{ij}$ denotes the covariance matrix of the datapoints. In addition we know that if $L/N_\tau \gg 1$, $\rho$ does not reproduce the datapoints within their errorbars, while if  $L/N_\tau \ll 1$ the spectrum will contain many unnatural structures arising from overfitting the noise in the data. Hence the most neutral reconstruction will satisfy $L/N_\tau = 1$, which we impose as constraint arising from prior knowledge. Hence our likelihood probability reads
\begin{align}
 P[D|\rho,I]={\rm exp}[-L - \gamma(L-N_\tau)^2]\label{PDrhoI}
\end{align}
where the limit $\gamma\to\infty$ is taken numerically. Note that maximizing this expression alone is still ill-defined, since the $N_\omega \gg N_\tau$ parameters $\rho_l$ are not yet uniquely fixed. 

Hence we continue by specifying the prior probability $P[\rho|I]$, which acts as a regulator and will allow us to select a unique Bayesian set of $\rho_l$'s. The MEM utilizes the Shanon-Jaynes entropy $S_{SJ}$ in its prior probability \cite{Jarrell:1996}, which is constructed from four axioms. Similarly we introduce our expression for $P[\rho|I]\propto {\rm exp}[S]$, after replacing two of these axioms. 

Prior information is incorporated explicitly through a function $m(\omega_l)=m_l$, which, by definition, is the correct spectral function in the absence of data \cite{Jarrell:1996}. It is usually obtained through previous reconstructions with less accurate data or from independent estimates. We begin the construction of the functional S with:

\paragraph{Axiom I: Subset independence} Let us consider two different subsets $\Omega_1$ and $\Omega_2$ along the frequency axis. If prior information imposes constraints on the spectrum $\rho$ within each of these subsets, then the result of the reconstruction should not depend on treating these domains separately or in a combined fashion $S[\Omega_1,m(\Omega_1)]+S[\Omega_2,m(\Omega_2)]=S[\Omega_1\cup\Omega_2,m(\Omega_1\cup\Omega_2)].$
This relation is satisfied if $S$ is written as an integral over frequencies
\beq
S\propto \int d\omega\; s(\rho(\omega), m(\omega), \omega) \label{S1}.
\eeq
While this axiom coincides with the one used in the MEM we continue by introducing a new:

\paragraph{Axiom II: Scale invariance} In general $\rho(\omega)$ does not have to be a probability distribution. Indeed, depending on the observable the spectrum is associated with, its scaling can differ from $1/\omega$. We hence require that the choice of units for $\rho$ and $m$ shall not change the result of the reconstruction. I.e. we must construct our prior probability using ratios of $\rho/m$ only
\beq
S=\tilde{\alpha} \int d\omega\;  s\Big(\rho(\omega)/m(\omega)\Big). \label{S2}
\eeq
Now that the integrand $s$ does not carry a dimension, we introduce the dimensionfull hyperparameter $\tilde{\alpha}$ to also make the argument of the exponential dimensionless. 

\paragraph{Axiom III: Smoothness of the reconstructed spectra} The only certain information about the spectral function is that it is a positive definite and smooth function. Hence we wish the prior functional to impose these traits on the reconstructed spectrum even if no further prior information is known. I.e. in the case of $m(\omega)=m_0$, a smooth spectrum shall be chosen independently of $m_0$.

The strategy towards this end relies on penalizing spectra which deviate between two adjacent frequencies $\omega_1$ and $\omega_2$. If changing the ratio $r_l=\rho_l/m_0$ at the two frequencies does not change the values of $D^\rho$ beyond the errorbars then $S$ should favor $r_1=r_2$. The penalty between the case where the same value exists at both frequencies $r_1=r_2=r$ and the case where they differ by a small amount $r_1=r(1+\epsilon)$, $r_2=r(1-\epsilon)$ hence has to be independent of r and symmetric in whether $r_1\gtrless r_2$:
\beq
2s(r)-s(r(1+\epsilon))-s(r(1-\epsilon))=\epsilon^2C_2.
\eeq
This is precisely the discretized expression for the differential equation $-r^2 s''(r)=C_2$, whose solution yields

\beq
S=\tilde{\alpha}\int d\omega\; \Big(C_0-C_1\frac{\rho}{m}+C_2\ln\left(\frac{\rho}{m}\right)\Big).
\eeq
The remaining axiom is identical to the MEM case, since it establishes the Bayesian meaning of $m(\omega)$. 
\paragraph{Axiom IV: Maximum at the prior} In the absence of data, $S$ must become maximal at $\rho=m$. Conventionally its value at this point is chosen to vanish
\beq 
S(r=1)=0, \quad S'(r=1)=0, \quad S''(r=1)<0.
\eeq
The two first conditions fix the constants $C_0,\ C_1$ and $C_2$ up to an overall constant, which we absorb into the hyperparameter $\alpha\propto\tilde{\alpha}$. The last condition forces $\alpha$ to be positive. Our final result hence reads
\beq
S=\alpha\int d\omega\; \Big(1-\frac{\rho}{m}+\ln\left(\frac{\rho}{m}\right)\Big).\label{Eq:Sfinal}
\eeq
This new prior distribution is strictly concave and exhibits the same quadratic behavior around the minimum $\rho=m$ as the Shanon-Jaynes entropy $S_{SJ}$. Hence the uniqueness of its maximum can be established analogously to the MEM \cite{Asakawa:2000tr}. In the case where $\rho_l,m_l \ll 1/\alpha$ or $\rho_l\ll m_l$, their contribution to $S$ and to the variation $\frac{\delta S}{\delta \rho}$ is not suppressed, thus we avoid the asymptotic flatness inherent in $S_{SJ}$. Note also that a closed expression for the normalization of $P[\rho|I]=P[\rho|\alpha,m]=N_S^{-1}{\rm exp}[S]$ is available $N_S=\prod_{i=1}^{N_\omega}{\rm exp}[\alpha\Delta\omega_i](\alpha\Delta\omega_i)^{-\alpha\Delta\omega_i} m_i \Gamma(\alpha\Delta\omega_i)$.

\begin{figure*}[t!]
\hspace{-0.3cm}\includegraphics[angle=-90,scale=0.17]{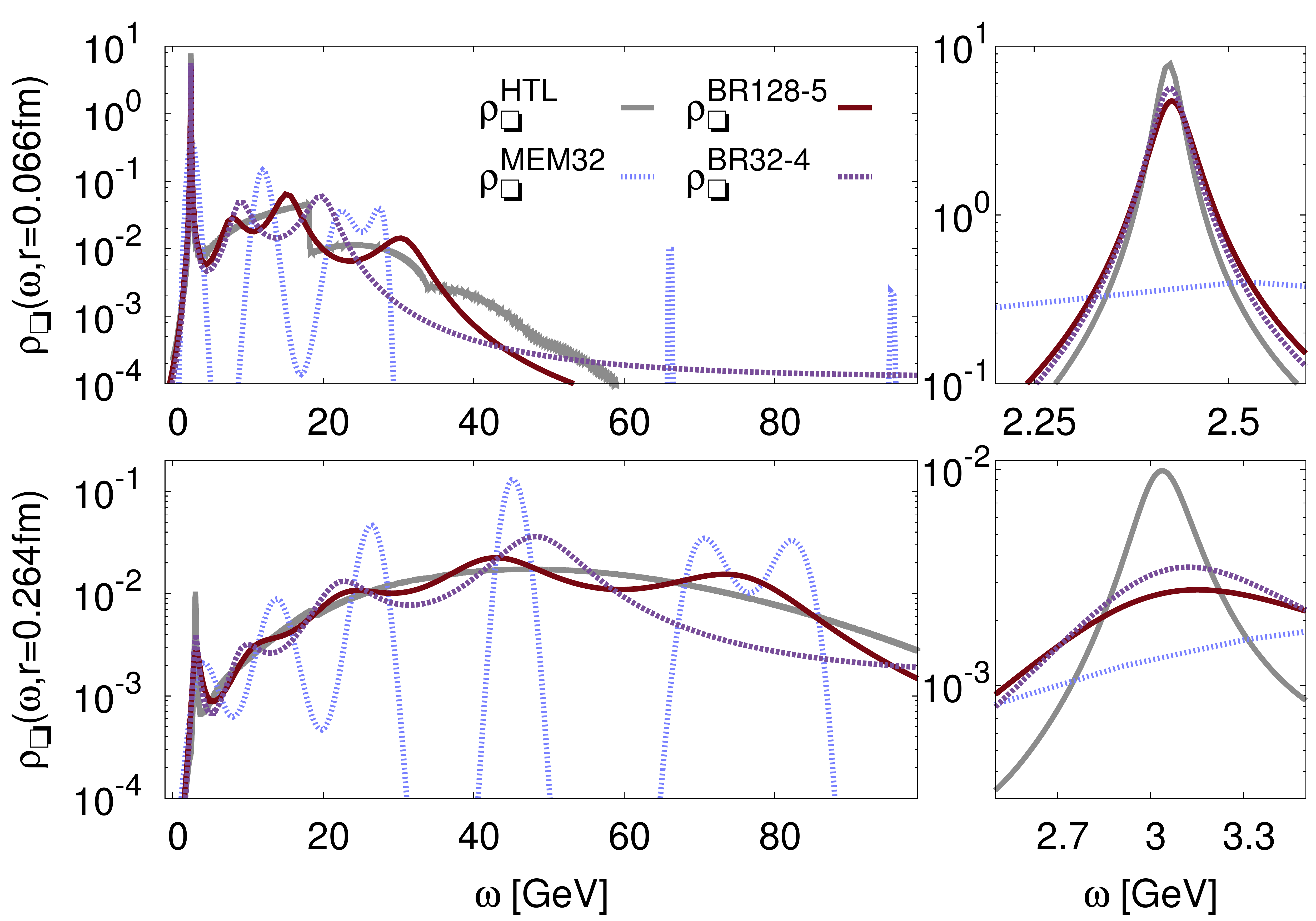}\hspace{0.8cm}\includegraphics[angle=-90,scale=0.17]{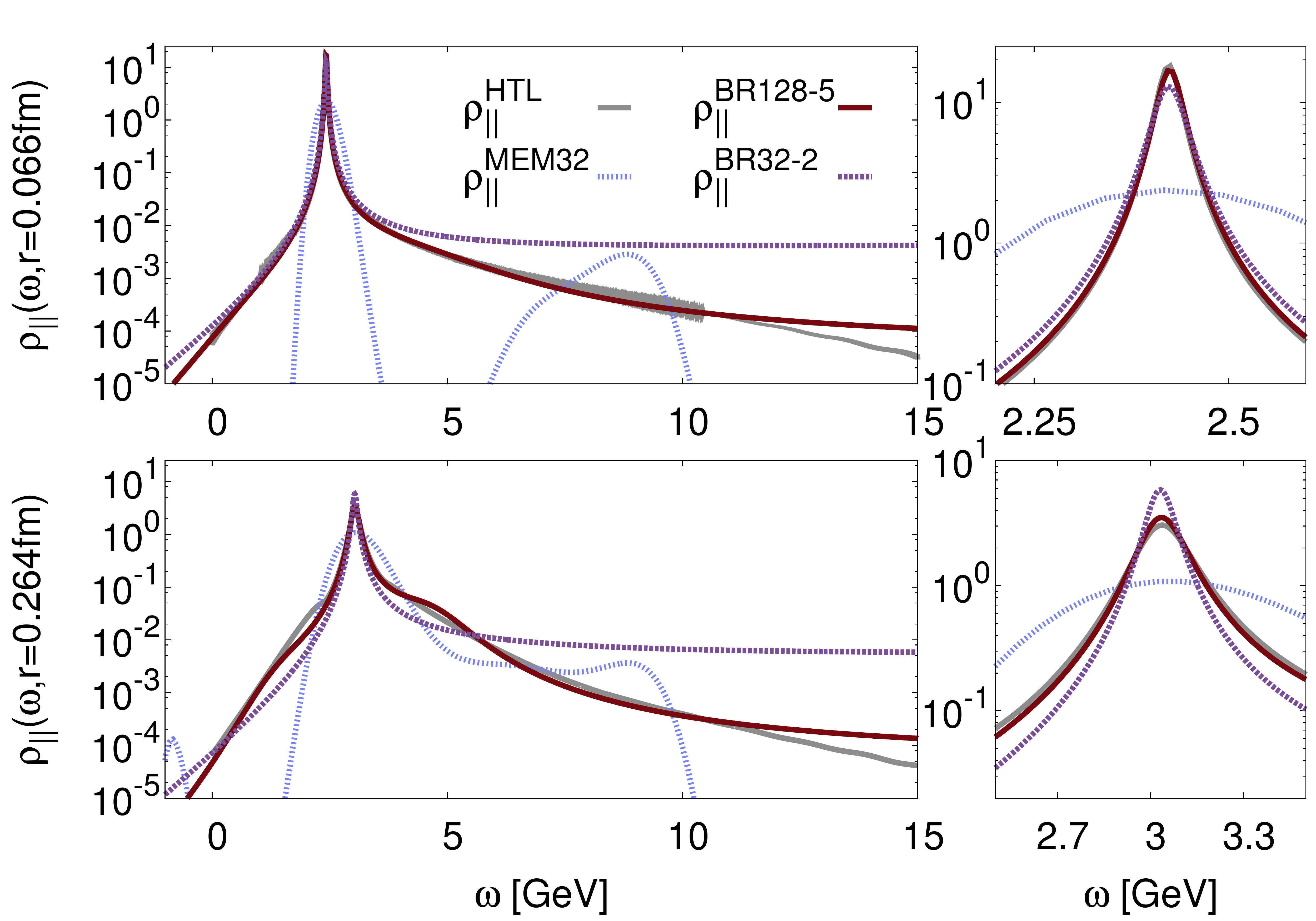} 
\caption{ Wilson Loop (left) and Wilson line correlator (right) spectra from HTL mock data \cite{Burnier:2013fca} at $r=0.066$fm (top) and $0.264$fm (bottom). As reference we show the HTL spectra $\rho^{\rm HTL}$ and the MEM $\rho^{\rm MEM32}$ \cite{Rothkopf:2011} with extended search space $N_{\rm ext}=80$ based on $N_\tau=32$ ideal datapoints. Both peak and background in $\rho^{\rm HTL}_\square$ are captured using $N_\tau=32$ and $\Delta D/D=10^{-4}$ (BR32-4), while the single peak in $\rho^{\rm HTL}_{||}$ requires only $\Delta D/D=10^{-2}$ (BR32-2). BR128-5 shows the result for a still realistic scenario of $N_\tau=128$ with mock error $\Delta D/D=10^{-5}$.}\label{Fig1} \vspace{-0.5cm}
\end{figure*}

By construction, our prior distribution contains the positive hyperparameter $\alpha$, which we need to treat in a Bayesian fashion. In the MEM, several spectra $\rho^\alpha$ for different values of $\alpha$, are reconstructed and ultimately averaged over \cite{Jarrell:1996}, weighted by the  values of the evidence $P[D|I]=P[D|\alpha,m]$. The calculation of the evidence relies however on a Gaussian approximation, whose validity is not guaranteed. 

Here we take a different route \cite{MacKay:1999}, and %self-consistently 
integrate out $\alpha$ from the joint probability distribution $P[\rho,D,\alpha,m]$. I.e. we take into account the influence of all possible prior distributions in the resulting posterior probability $P[\rho |D,m]$, on which we base the reconstruction. 
Starting from the multiplication law for
\bea
P[\rho,D,\alpha,m]&=&P[D|\rho,\alpha,m]P[\rho |\alpha,m]P[\alpha,m]\\
			     &=&P[\alpha |\rho,D,m]P[\rho |D,m]P[D,m],\notag
\eea
we integrate both r.h.s with respect to $\alpha$. Assuming no knowledge on the hyperparameter $(P[\alpha]=1)$ we set out to find an expression for the $\alpha$ independent 
\bea
P[\rho |D,m]=\frac{P[D|\rho,I]}{P[D|m]}\int d\alpha P[\rho | \alpha,m].\label{Eq:IntegrAlpha}
\eea
In the above expression $P[D|\rho,I]$ is given by Eq.\eqref{PDrhoI} and $P[D|m]$ is an irrelevant constant. For large values of S, we approximate the integral over $\alpha$ through a next-to-leading order resummation of logarithms, while for small S a numerical evaluation is possible. 

After integration, no dependence on $\alpha$ remains. The presence of $m$ is not problematic, for its constant values do not influence the reconstruction result if $\rho$ is properly normalized. Hence $P[\rho|D,m]$ does not contain any meaningful external parameters and we can proceed to find its maximum numerically. To this end we deploy the quasi-newton LBFGS algorithm, which allows us to approach $\delta P[\rho|D,m]/\delta \rho=0$ by varying each of the $N_\omega$ parameters $\rho_l$ individually. An inversion of the Hessian matrix at intermediate steps is not required, which leads to a significant reduction in computational cost compared to the usual Levenberg-Marquardt approach. Note that in contrast to the MEM with $S_{SJ}$, now without asymptotically flat directions, we successfully locate the global extremum of $P[\rho|D,m]$ and do not need to stop the algorithm at an artificial cutoff in step size. %More details will be given in a forthcoming publication.

One area of application is the static potential between two heavy quarks, which is related to the spectral structure of the rectangular Wilson loop $W_\square(r,\tau)$ and possibly the Wilson line correlator in Coulomb gauge $W_{||}(r,\tau)$ \cite{Rothkopf:2011db}.  It has been shown that the extraction of such spectra poses a severe challenge to the MEM \cite{Burnier:2013fca}. 

Here we benchmark our approach through reconstruction of known spectra from cutoff regularized ($\Lambda=5\pi$) Euclidean correlators, calculated in hard thermal loop (HTL) resummed perturbation theory \cite{Burnier:2013fca}. Subsequently we attempt to extract from them the known HTL inter-quark potential \cite{Laine:2007qy} by fitting the position and width of the lowest lying peak \cite{Rothkopf:2011db}. The ideal HTL data points are perturbed by Gaussian noise with variance $\sigma_i^2= (\eta D_i)^2$, leading to constant relative errors $\Delta D/D=\eta$. In preparation for lattice QCD, we assume no prior knowledge on the spectrum and supply a constant prior $m(\omega)=1$.

In Fig.\ref{Fig1} we present reconstructions ($N_\omega=1000$, $\omega\in[-126,189]{\rm GeV}$ (left) , $N_\omega=1200$, $\omega\in[-15.7,15.7]{\rm GeV}$ (right)) from mock data at $T=2.33T_C$ for qualitative comparison. The reference MEM \cite{Rothkopf:2011} ($\rho^{\rm MEM32}$) based on $N_\tau=32$ ideal datapoints \cite{Burnier:2013fca} fails to reproduce even the Lorentzian shape of the lowest peak in the HTL spectrum ($\rho^{\rm HTL}$). $\rho^{\rm HTL}_\square$ contains a peak and a large background, both of which our method is able to capture with $N_\tau=32$ at $\Delta D/D=10^{-4}$. In $\rho^{\rm HTL}_{||}$ a single peak is dominant for which $N_\tau=32$ at $\Delta D/D=10^{-2}$ suffices. To showcase possible improvements for future lattice QCD studies, we also present the results for $N_\tau=128$ datapoints and $\Delta D/D=10^{-5}$ (BR128-5).

\begin{figure}[t!]
\includegraphics[angle=-90,scale=0.165]{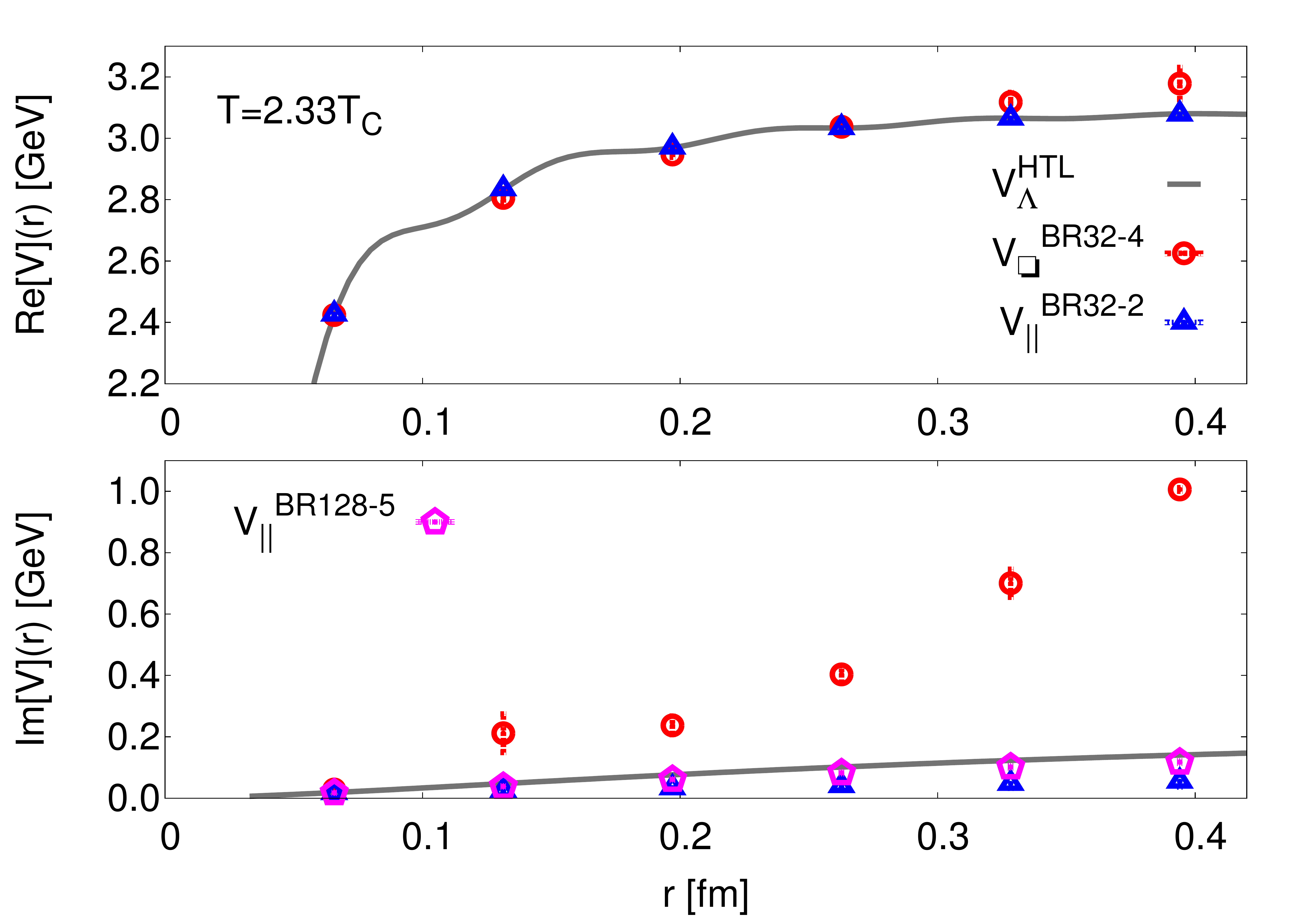}
\caption{Reconstruction of the HTL ($T=2.33T_C$) potential (solid line) based on the inferred spectra from the HTL Wilson Loop $V_\square(r)$ (circle) and the HTL Wilson line correlator $V_{||}(r)$ (triangle). (top) ${\rm Re}[V](r)$ is reproduced with sub $10\%$ and sub $1\%$ deviation from $N_\tau=32$ data points with $\Delta D/D=10^{-4}$ and $10^{-2}$ respectively. On the other hand ${\rm Im}[V](r)$ (bottom) requires at least $N_\tau=128$ with $\Delta D/D=10^{-5}$ for $20\%$ accuracy (pentagon).}\label{Fig2} \vspace{-0.4cm}
\end{figure}

As quantitative check, we reproduce the known HTL inter-quark potential from the lowest spectral peak. Our method hence needs to yield the correct position $({\rm Re}[V])$ and width $({\rm Im}[V])$ of a skewed Lorentzian. In the top panel of Fig.\ref{Fig2} we show the real part (solid line) obtained from the Wilson Loop $V_\square(r)$ (circle) and Wilson line correlators $V_{||}(r)$ (triangle). Error bars are estimated from three reconstructions with different mock noise of equal strength. With $N_\tau=32$ the correct real part is reproduced with $10\%$ and $1\%$ accuracy respectively; especially the strong artificial rise in ${\rm Re}[V_\square]$ observed in the MEM in Ref.\cite{Burnier:2013fca} is absent. 

With our method it is also possible to reproduce the width from the Wilson line correlators. To this end we utilize $N_\tau=128$ and $\Delta D/D=10^{-5}$, which is still realistic in quenched lattice QCD. The resulting reconstruction of ${\rm Im}[V](r)$ with sub $20\%$ deviation is shown in the lower panel of Fig.\ref{Fig2} (pentagon). (For $N_\tau=128$ we only show ${\rm Im}[V_{||}](r)$, since the background from cusp divergences in $\rho_\square$ \cite{Burnier:2013fca} will necessitate even better data.)

\begin{figure}[t]
\includegraphics[angle=-90,scale=0.165]{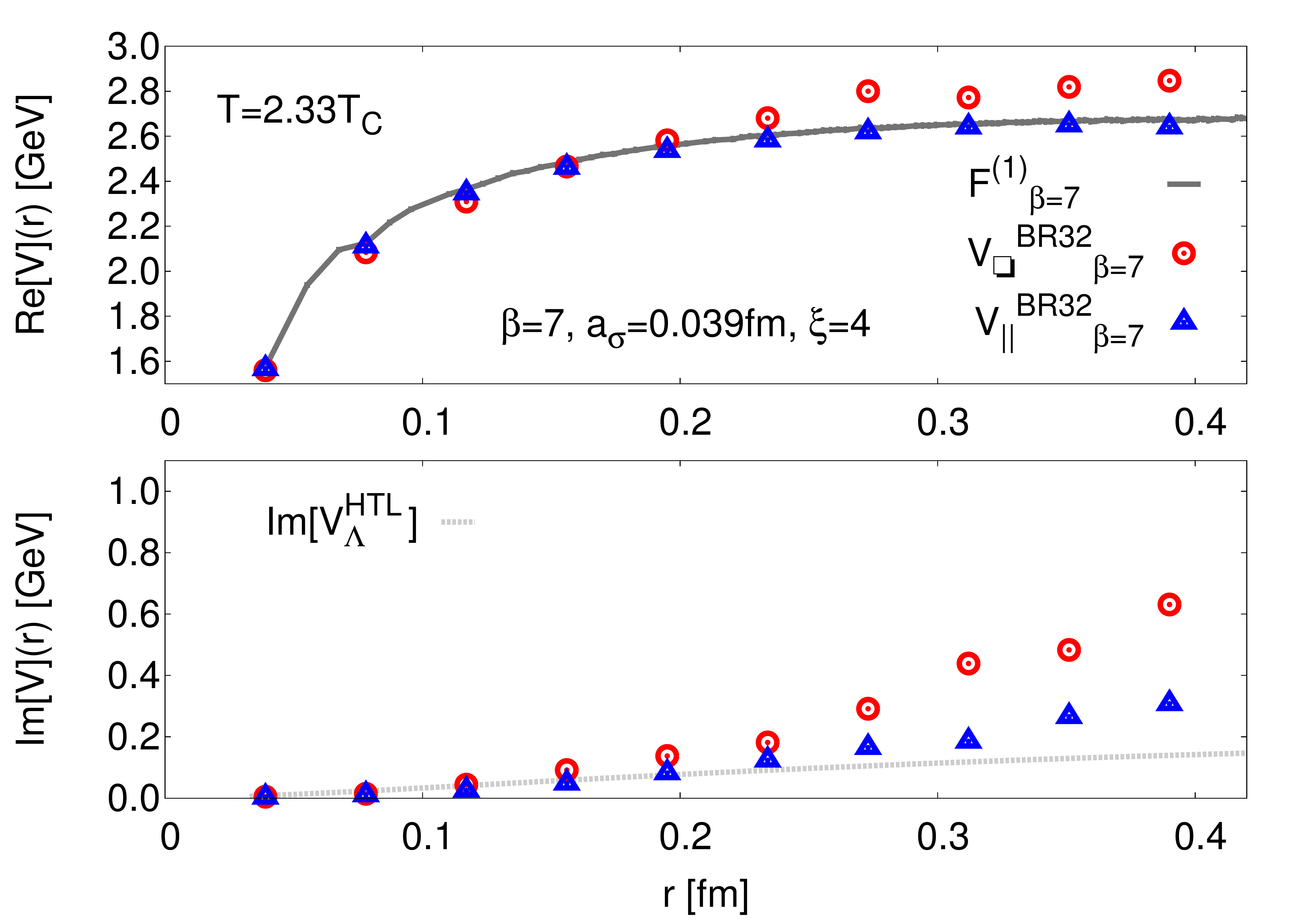}
\caption{Improved estimate for the heavy quark potential from quenched lattice QCD Wilson Loops $V_\square(r)$ (circle) and Wilson line correlators $V_{||}(r)$ (triangle). (top) At the small values of $r<0.4$fm shown, ${\rm Re}[V](r)$ appears to lie close to the Coulomb gauge color singlet free energies $F^{(1)}$ (solid line). (bottom) At this $T=2.33T_C$ we find an ${\rm Im}[V](r)$ that is of the same order as in HTL perturbation theory (dashed line)}\label{Fig3}\vspace{-0.4cm}
\end{figure}

With these limitations in mind, we apply our method to the  Wilson Loop and Wilson lines in quenched lattice QCD \cite{Rothkopf:2011db} at $T=2.33T_C$ ($N_\tau=32$, $\beta=7$, $a_\sigma=0.0039$fm, $\xi=4$). The improved estimate in Fig.\ref{Fig3} for both ${\rm Re}[V_\square]$ and ${\rm Re}[V_{||}]$ shows that, as expected at the small distances treated here \cite{Burnier:2009bk}, their values lie close to the color singlet free energies in Coulomb gauge $F^{(1)}$. While we do not expect the width, i.e. ${\rm Im}[V]$ (bottom), to be captured reliably yet at $N_\tau=32$, it is interesting to note that its values appear to be of the same order of magnitude as in the HTL calculation at this temperature.

We have introduced a novel Bayesian approach to spectral function reconstruction. It cures the conceptual and practical issues affecting the MEM by introducing an improved dimensionless prior distribution devoid of asymptotically flat directions in Eq.\eqref{Eq:Sfinal}. In the case of a constant prior function $m(\omega)=m_0$ and normalization of $\rho$, no external parameter needs to be adjusted, since we integrate out explicitly the hyperparameter $\alpha$ as shown in Eq.\eqref{Eq:IntegrAlpha}. Combined with the LBFGS optimizer algorithm, which varies each of the individual $N_\omega$ parameters $\rho_l$, we achieve a significant improvement in the reconstruction of spectra as demonstrated in Fig.\ref{Fig1} and Fig.\ref{Fig2}. Hence we look forward to further applications in lattice QCD (Fig.\ref{Fig3}) and beyond.

The authors thank T.~Hatsuda, S.~Sasaki, O.~Kaczmarek, S.Y.~Kim, P.~Petreczky and H.T.~Ding for fruitful discussions on the MEM, C.A.~Rothkopf for insight on Bayesian inference and the DFG-Heisenberg group of Y.~Schr\"oder at Bielefeld Univ. for computing resources. This work was partly supported by the Swiss National Science Foundation (SNF) under grant 200021-140234.

\end{document}